\documentclass[]{spie} 
\usepackage[]{graphicx}
\usepackage{amsmath}
\usepackage{booktabs}
\usepackage{aas_macros}
\usepackage{cite}
\DeclareGraphicsExtensions{%
    .png,.PNG,%
    .pdf,.PDF,%
    .jpg,.mps,.jpeg,.jbig2,.jb2,.JPG,.JPEG,.JBIG2,.JB2}

\title{Development of a scalable generic platform for adaptive optics real time control} 

\author{Avinash Surendran\supit{a}, Mahesh P. Burse\supit{b}, A. N. Ramaprakash\supit{b}, Padmakar Parihar\supit{a}
\skiplinehalf
\supit{a}Indian Institute of Astrophysics, Koramangala 2nd Block, Bangalore - 560034, India \\
\supit{b}Inter-University Centre for Astronomy and Astrophysics, No. 4, Ganeshkhind, Pune University Campus, Pune - 411007, India
}

\authorinfo{Further author information: (Send correspondence to A.S.)\\A.S.: E-mail: asurendran@iiap.res.in}
 
\begin{document} 
  \maketitle 

\begin{abstract}
The main objective of the present project is to explore the viability of an adaptive optics control system based exclusively on Field Programmable Gate Arrays (FPGAs), making strong use of their parallel processing capability. In an Adaptive Optics (AO) system, the generation of the Deformable Mirror (DM) control voltages from the Wavefront Sensor (WFS) measurements is usually through the multiplication of the wavefront slopes with a predetermined reconstructor matrix. The ability to access several hundred hard multipliers and memories concurrently in an FPGA allows performance far beyond that of a modern CPU or GPU for tasks with a well defined structure such as Adaptive Optics control. The target of the current project is to generate a signal for a real time wavefront correction, from the signals coming from a Wavefront Sensor, wherein the system would be flexible to accommodate all the current Wavefront Sensing techniques and also the different methods which are used for wavefront compensation. The system should also accommodate for different data transmission protocols (like Ethernet, USB, IEEE 1394 etc.) for transmitting data to and from the FPGA device, thus providing a more flexible platform for Adaptive Optics control. Preliminary simulation results for the formulation of the platform, and a design of a fully scalable slope computer is presented.
\end{abstract}


\section{INTRODUCTION}
\label{sec:intro}

The technology of AO is used to correct the fast-changing blurring caused by the Earth's atmosphere in real time. The most distinctive components of an adaptive optics system are: the DM which dynamically corrects for the optical aberrations of the incoming signal; the WFS which measures the shape of the instantaneous wavefront distortions due to atmospheric turbulence; and the control system which interfaces both the DM and WFS to correct for atmospheric turbulence at a high frequency. The technique is now helping astronomers attain near-diffraction limited resolution images at IR wavelengths with the largest existing telescopes, without the cost and difficulty associated with sending them to space. The landscape of adaptive optics has reached a level which demands abstraction of its core functions, where a large part of the computational algorithms used for AO control are common irrespective of the telescope that they are used on. With the advent of extremely large telescopes (30 m -- 100 m), we find the need for incorporating AO as an indispensable part of these telescopes. Hence, there needs to be a way to implement the same with a minimal development overhead and with a great flexibility of wavefront sensing and correction resources. The Thirty Meter Telescope would pave the way as the first telescope designed with AO as an integral system element \cite{2014SPIE.9148E..10H}.

The most computationally intensive processes for an AO system are slope computation and phase reconstruction. For a standard least-squares reconstructor algorithm, the computational complexity scales as $\rm{FLOPS}\sim\rm{D}^4$ most of which is contributed by full matrix-multiply operations\cite{2002SPIE.4494..215G}. For e.g., extreme-AO on TMT would require $10^{5}$ times the processing power of the current Keck AO system\cite{2006PASP..118..297W}. This level of parallelism and high performance computing requirement of AO have conventionally been implemented with GPUs \cite{2012SPIE.8447E..23W} (Graphical Processing Units), and multi-core CPUs \cite{2014SPIE.9148E..10H}. FPGAs are semiconductor devices that are based around a matrix of configurable logic blocks (units of logic gates and flip-flops) connected via programmable interconnects, and can be reprogrammed to implement any digital design. The FPGA leads in the ability to easily interface high speed I/O and a reduced memory latency but it is slower at floating point operations compared to the GPU\cite{minhas2014gpu}. We need to explore FPGAs as a viable alternative to GPUs and multi-core CPUs in providing increased system performance at lower costs and a better power efficiency for industrial adaptive optics systems. The use of FPGAs for the next generation of high-end ground based telescope AO controllers has already been explored by a number of researchers\cite{2012MNRAS.424.1483B,ramos2008fpga,zhang2012high}, but a truly scalable and generic platform still doesn't exist.

In Section~\ref{sec:RTC}, we show the requirements of the Real Time Controller (RTC) and the rationale behind choosing them. In Section~\ref{sec:WPU}, we present the implementation of a fully scalable wavefront processing unit (WPU). In Section~\ref{sec:result}, we present the results including the logic resources and the timing summary.

\section{RTC REQUIREMENTS AND CHALLENGES}
\label{sec:RTC}
The key requirements of the RTC are as follows:
\begin{enumerate}
\item To create a generic platform for adaptive optics, which can incorporate current and future wavefront sensing and correction techniques through abstraction of the wavefront sensing and correction geometries, as described by Southwell.\cite{southwell1980wave}.
\item To incorporate the flexibility of different I/O protocols for wavefront sensing, correction and telemetry into the FPGA module.
\item To make it scalable to fulfil short term requirements of flexibility in telescope mirror size, atmospheric conditions, the frame rate of acquisition, and the number and nature of guide stars used.
\end{enumerate}

\subsection{Error Budgets}
We have set a target Strehl ratio of 0.5 at a wavelength of 1 $\mu$m owing to the extreme AO capability that would be required for the next generation of Extra Large Telescopes (ELTs) and the limitations of error budgets that can be achieved with current technology. A Strehl ratio of 0.5 corresponds to a total RMS wavefront error (WFE) of about 130 nm at a wavelength of 1 $\mu$m, according to Marechal approximation \cite{2002SPIE.4494..215G}. The degrees of freedom (DOF) of an AO system is determined by the spatial resolution of wavefront detection and correction, and thus directly contributes to the fitting error\cite{roddier1999adaptive}. A 10,000 DOF system translates to a fitting error of 64 nm for a \mbox{30 m} telescope, which is an acceptable limit for achieving the required Strehl ratio.

\subsection{Interface Control Requirements}
\label{subsec:ICRD}
The function of the WPU is mainly pixel acquisition and slope computation. It should also have the capability for preliminary image processing of the RAW CCD frame including (but not limited to) dark subtraction, flat fielding and background correction. The WPU is initially assumed to be able to interface to a Shack-Hartmann (SH) sensor which uses Fried geometry\cite{fried1977least}. The error budget is used to derive the WPU interfacing requirements which are enumerated as follows:
\begin{enumerate}
\item It should be able to interface four WFS CCDs of size 512 x 512 pixels at 16-bit/pixel, through a maximum of a total of 16 fibre channels.
\item It should be able to read out an entire CCD frame in 0.5 ms.
\item Pixels per subaperture should be flexible anywhere between 2 x 2 to 8 x 8 pixels.
\end{enumerate}
The above requirements translate to a memory bandwidth of 256 MB/s for a single fiber channel, which in turn leads to a requirement of 1 GB/s for a single CCD through 4 fibre channels and 4 GB/s for 4 CCDs through 16 fibre channels.

\subsection{AO Reconstructor Requirements}
Zonal reconstruction using a conventional Matrix Vector Multiply (MVM) operation will be initially adopted for its simplicity and scalability\cite{herrmann1980least}. If we assume a sensor size of 512 x 512 pixels with each subaperture consisting of 4 x 4 pixels, the number of slopes along a single spatial dimension is 128 and the number of WF slopes generated by a single sensor would be 32,768. For a single conjugate AO (SCAO) system, the size of the reconstruction matrix would be \mbox{32,768-by-16,641}, where 16,641 is the number of actuators of the DM without any form of phase interpolation. If we assume that each element of the reconstruction matrix can be represented in 2 byte fixed-point format, it would occupy 1 GB. For a Multi-Conjugate AO (MCAO) system with 4 sensors, it would scale up to a 131,072-by-16,641 matrix which would occupy 4 GB. An AO control loop frequency of 800 Hz would require a computational power of 3.48 TFlops for AO reconstruction. The memory bandwidth would heavily depend on the hardware used and the method of implementation.

\section{Implementation of a Scalable WPU}
\label{sec:WPU}
\subsection{MATLAB AO Simulator}
A basic scalable SCAO simulator was made using MATLAB wherein one could change the number of subapertures and the pixels per subaperture of a basic SH (Shack-Hartmann) sensor in a Fried geometry configuration. It is coupled with a test-bench where the Von-Karmann phase screen could be generated with atmospheric input parameters including but not limited to Fried parameter, aperture diameter, the inner and outer scales of turbulence etc. The phase screen generation can be scaled upto very large aperture diameters and is based on the work by Sedmak\cite{sedmak2004implementation}. Testing of all slope computation and AO reconstruction algorithms is done on this simulator before implementation on an FPGA.

\subsection{Test Platform}
The Xilinx VC-709 Connectivity Board\cite{vc709manual} is chosen as the test platform mainly owing to the powerful \mbox{Virtex-7} XC7VX690T FPGA coupled with the ability of flexible and fast hardware interfacing options like multiple small form-factor pluggable (SFP) transceiver ports, Peripheral Component Interconnect Express (PCIe) 3.0 x8, FPGA Mezzanine Connector (FMC) and 2 banks of 4 GB DDR3-1866 memory modules. 

\subsection{Objectives}
The first step is to design a fast pixel acquisition platform coupled with a pipelined and fully scalable slope computer for an SH sensor. The WPU is designed based on the following assumptions:
\begin{enumerate}
\item The platform should support a full-frame readout time of 0.5 ms for a 4-quadrant 512 x 512 pixel CCD (from Section~\ref{subsec:ICRD}), which translates to a pixel clock period of 7.629 ns, or a clock frequency of \mbox{131.072 MHz}.
\item The architecture should be scalable with respect to the number of subapertures and the pixels per subaperture, and be modular for testing different slope computation algorithms.
\item A similarly scalable VHDL test-bench should be linked with MATLAB AO simulator for quick troubleshooting and validation.
\end{enumerate}

\subsection{Implementation}
\subsubsection{Dataflow}
\begin{figure}[ht]
\begin{center}
\includegraphics[width=0.9\textwidth]{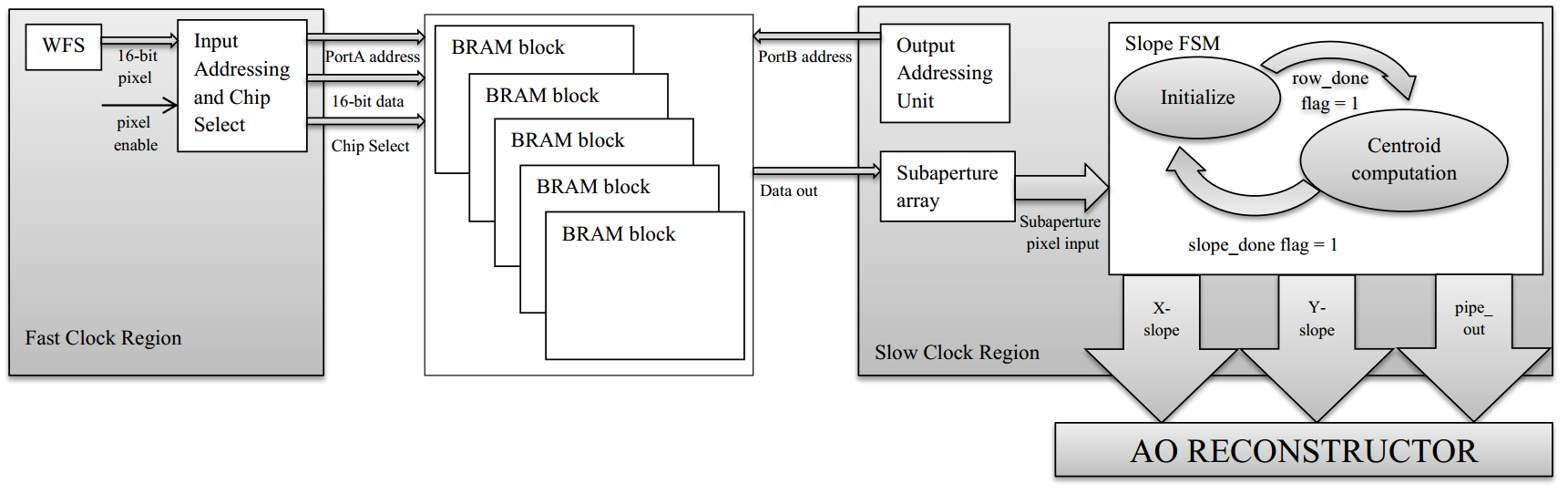}
\caption{WPU Dataflow}
\label{fig:DF}
\end{center}
\end{figure}

A combination of Dual-port Block RAM (BRAM), Finite State Machine (FSM) design and clock management techniques are used to implement a fast, scalable and efficient WPU. For slope computation, we use a conventional Centre of Gravity (CoG) algorithm\cite{roddier1999adaptive} for the sake of demonstration. Division essentially consists of repeated subtraction, and hence the CoG takes a large datapath delay. The design is divided into two clock domains, one for fast pixel acquisition which runs at the CCD pixel clock frequency and the other for slope computation which runs at $\frac{1}{16}$th the frequency of the pixel clock. The BRAM is used to isolate the 2 clock domains as shown in Figure~\ref{fig:DF}. The `Input Addressing and Chip Select' unit identifies the pixel location in the CCD and generates the address and BRAM block to be selected where the pixel is to be stored. The pixel acquisition unit can work at a maximum frequency of 131.072 MHz.

The state of any FSM is decided by its current state and any input variables associated with it. The slope FSM starts at the `Initialize' state and waits till enough pixel data is available in the BRAM for slope computation. When the last pixel of a particular row of subapertures is read out into the BRAM, enough information is available to commence slope computation and the state of the FSM changes to `Centroid Computation'. The number of x-slopes and y-slopes which are available at the output at every clock cycle depends on the value of a user-defined parameter called `iter'. A larger `iter' translates to a faster slope computation at the expense of the FPGA logic resources. `pipe$\_$out' gives the information of the number of slopes that have already been computed, to the AO reconstructor. Both the pixel acquisition and slope computation occur in parallel without one affecting the performance of the other. The FSM can be operated at a maximum frequency of around 10 MHz.

\subsubsection{BRAM configuration example}

\begin{figure}[ht]
\begin{center}
\includegraphics[width=0.75\textwidth]{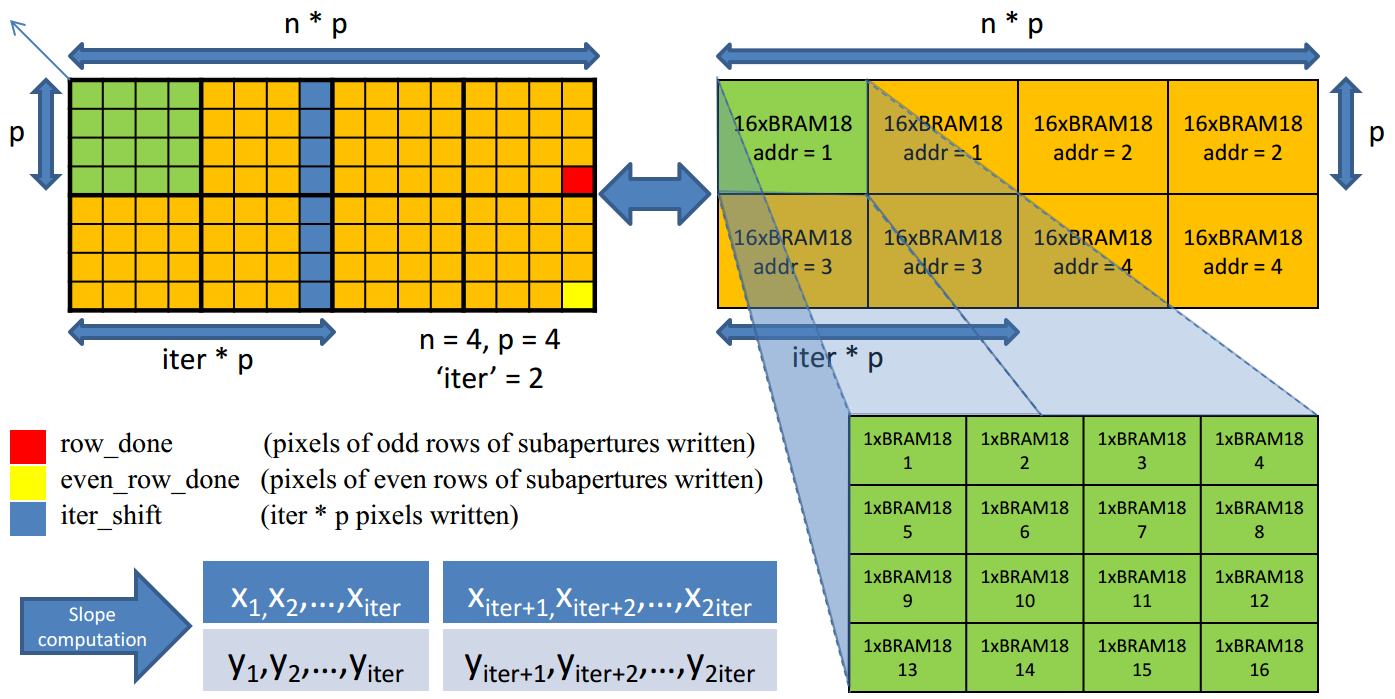}
\caption{BRAM Configuration example. n is the number of subapertures, p is the pixels per subaperture and iter is the number of slopes which are computed per clock cycle.}
\label{fig:config}
\end{center}
\end{figure}
BRAM36 is a BRAM primitive present in the Virtex-7 series of FPGAs and can be divided into two BRAM18 primitives. It has a 32-bit datawidth with one block being able to store 1K datawords. In the example shown in Figure~\ref{fig:config}, the number of subapertures along a row (n) is set to 4, the pixels per subaperture along a row (p) is set to 4 and the slopes to be computed per clock cycle (iter) is set to 2. The pixels of every `iter' subapertures are stored in different BRAM blocks as all the associated pixels need to be concurrently accessed for slope computation in a single clock cycle. The pixels associated with the first two (as $\rm{iter}=2$) subapertures are stored in the first address of different BRAMs for concurrent access. The pixels of the third and fourth subaperture are stored in the second address of the same set of BRAM blocks, and the pattern continues for two rows of subapertures. When the slope is being computed for one row of subapertures, the pixels of the next row of subapertures will be written into the BRAM. Hence, memory access for the pixels of two rows of subapertures are required at a time as shown in Figure~\ref{fig:config}. The flag, `row$\_$done' signifies the end of each row of subapertures and triggers the FSM to start centroid computation. `even$\_$row$\_$done' signifies the completion of readout of the pixels of an even number of subapertures and resets the addressing logic.

\section{Results and Conclusion}
\label{sec:result}
The final slope computer is fully scalable with respect to the number of subapertures (n), pixels per subaperture (p) and the slopes to be computed per clock cycle of the FSM (iter). The architecture is modular, wherein the CoG algorithm can be replaced by any other slope computation algorithm with negligible programming effort. The slopes are obtained in a binary fixed-point format with 8 decimal places. The difference in accuracy when the decimal accuracy was increased to 16 decimal places was found to be negligible in AO reconstruction, using the MATLAB AO simulator. The simulated output of the WPU was compared with the results from the MATLAB AO simulator and they agreed to the 8th binary decimal place, as expected. From Table~\ref{table:single} and Table~\ref{table:quad}, we can see that the logic resource usage is independent of the number of subapertures (n). It only depends on the pixels per subaperture (p) and the slopes to be computed per clock cycle (iter). We can also conclude that the WPU design is meeting the logic resource limits and timing requirements, and is feasible to be implemented on the Xilinx VC-709 Connectivity Board. The same platform can also be implemented on cheaper Xilinx FPGAs with minimal changes, and can be used in small scale AO for telescopes having an aperture diameter of 2 -- 4 m.

\begin{table}[h]
\begin{tabular}{@{}lllllllll@{}}
\toprule
n  & p & iter & Slice Registers & Slice LUT & BRAM36 	   & WNS (ns) & DSP & FPGA utilization \\ \midrule
32 & 4 & 16   & 650             & 40010     & 128          & 1.078    & 128 & 11\%                  \\
   &   & 8    & 345             & 19837     & 64           & 1.3      & 64  & 5.40\%                \\
   &   & 4    & 194             & 9423      & 32           & 2.272    & 32  & 2.60\%                \\
32 & 8 & 16   & 1510            & 98378     & 512          & 0.626    & 128 & 27\%                  \\
   &   & 8    & 797             & 48852     & 256          & 0.821    & 64  & 13.80\%               \\
   &   & 4    & 420             & 24718     & 128          & 1.279    & 32  & 7\%                   \\
64 & 4 & 32   & 1269            & 79698     & 256          & 0.336    & 256 & 22\%                  \\
   &   & 16   & 658             & 39903     & 128          & 1.198    & 128 & 11\%                  \\
   &   & 8    & 349             & 19942     & 64           & 0.791    & 64  & 5.50\%                \\
64 & 8 & 32   & 2985            & 195959    & 1024         & 0.24     & 256 & 55\%                  \\
   &   & 16   & 1541            & 97978     & 512          & 0.641    & 128 & 28\%                  \\
   &   & 8    & 816             & 49005     & 256          & 0.896    & 64  & 13.80\%               \\ \midrule
\multicolumn{3}{|c|}{Total FPGA resources} & 866400 & 433200 & 1470 & & 3600 & \\ \bottomrule
\end{tabular}
\caption{FPGA resource usage and timing summary in terms of Worst Negative Slack (WNS) for the WPU interfaced to a single channel CCD. n is the number of subapertures, p is the pixels per subaperture and iter is the number of slopes which are computed per clock cycle.}
\label{table:single}
\end{table}

\begin{table}[h]
\begin{tabular}{@{}llllllll@{}}
\toprule
n  & p & iter & Slice Registers & Slice LUT & BRAM36 & DSP & FPGA Utilization \\ \midrule
64 & 4 & 16   & 2632            & 177475    & 512  & 512 & 45\%      \\
   &   & 8    & 1396            & 80172     & 256  & 256 & 22.20\%   \\ \midrule
\multicolumn{3}{|c|}{Total FPGA resources} & 866400 & 433200 & 1470 & 3600 & \\ \bottomrule
\end{tabular}
\caption{FPGA resource usage for the WPU interfaced to a four channel CCD. n is the number of subapertures, p is the pixels per subaperture and iter is the number of slopes which are computed per clock cycle.}
\label{table:quad}
\end{table}

\bibliography{bib}   
\bibliographystyle{spiebib}

\end{document}